%% file: matrix_PIT.tex
\documentclass[11pt]{article}

\usepackage{pictexwd,dcpic}

\input{preamble}
\usepackage{fullpage}

\newcommand{\mat}[1]{\insquar{\begin{array}{cc}#1\end{array}}}
\newcommand{\matr}[1]{\insquar{\begin{array}{ccc}#1\end{array}}}
\begin{document}

\title {The Power of Depth $2$ Circuits over Algebras}

\author{
Chandan Saha\thanks{Indian Institute of Technology, Kanpur 208016, India. {\tt Email - csaha@cse.iitk.ac.in}}
  \and
  Ramprasad Saptharishi\thanks{Chennai Mathematical Institute, Chennai
    603103, India. Supported by MSR India PhD Fellowship. {\tt Email - ramprasad@cmi.ac.in}}
  \and
  Nitin Saxena\thanks{Hausdorff Center for Mathematics, Bonn 53115,
    Germany. {\tt Email - ns@hcm.uni-bonn.de}}
}

\date{}

\maketitle
\begin{abstract}
We study the problem of polynomial identity testing (PIT) for depth
$2$ arithmetic circuits over matrix algebra.  We show that identity
testing of depth $3$ ($\Sigma \Pi \Sigma$) arithmetic circuits over a
field $\F$ is polynomial time equivalent to identity testing of depth
$2$ ($\Pi \Sigma$) arithmetic circuits over
$\mathsf{U}_2(\mathbb{F})$, the algebra of upper-triangular $2\times
2$ matrices with entries from $\F$. Such a connection is a bit
surprising since we also show that, as computational models, $\Pi
\Sigma$ circuits over $\mathsf{U}_2(\mathbb{F})$ are strictly `weaker'
than $\Sigma \Pi \Sigma$ circuits over $\mathbb{F}$.

The equivalence further shows that PIT of depth $3$ arithmetic
circuits reduces to PIT of width-$2$ planar commutative
\emph{Algebraic Branching Programs}(ABP). Thus, identity testing for
commutative ABPs is interesting even in the case of width-$2$.

Further, we give a deterministic polynomial time identity testing
algorithm for a $\Pi \Sigma$ circuit over any constant dimensional
commutative algebra over $\mathbb{F}$. While over 
commutative algebras of polynomial dimension, identity testing is 
at least as hard as that of $\Sigma \Pi \Sigma$ circuits 
over $\mathbb{F}$.

\end{abstract}

\section{Introduction} \label{sec:intro}
Polynomial identity testing (PIT) is a fundamental problem in
theoretical computer science. Over the last decade this problem has
drawn significant attention from many leading researchers owing to its
role in designing efficient algorithms and in proving circuit lower
bounds. Identity testing is the following problem:

\begin{problem}
Given an arithmetic circuit $C$ with input variables $x_1, \ldots,
x_n$ and constants taken from a field $\F$, check if the polynomial
computed by $C$ is identically zero.
\end{problem}

Besides being a natural problem in algebraic computation, identity
testing appears in important complexity theory results such as,
$\mathsf{IP} = \mathsf{PSPACE}$ \cite{Shamir90} and the PCP theorem
\cite{ALMSS98}. It also plays a promising role in proving
super-polynomial circuit lower bound for permanent
\cite{KI03,A05}. Moreover, algorithms for problems like primality
testing \cite{AKS04}, graph matching \cite{L79} and multivariate
polynomial interpolation \cite{CDGK91} also involve identity testing.

The first randomized polynomial time algorithm for identity testing
was given by Schwartz and Zippel \cite{S80,Z79}. Several other
efficient randomized algorithms \cite{CK97,LV98,AB99,KS01} came up
subsequently, resulting in a significant improvement in the number of
random bits used.  However, despite many attempts a deterministic
polynomial time algorithm has remained elusive. Nevertheless,
important progress has been made both in the designing of
deterministic algorithms for special circuits, and in the
understanding of why a general deterministic solution could be hard to
get. \\

Without loss of generality, we can assume that a circuit $C$ has
alternate layers of addition and multiplication gates. A layer of
addition gates is denoted by $\Sigma$ and that of multiplication gates
is denoted by $\Pi$. Kayal and Saxena \cite{KS07} gave a deterministic
polynomial time identity testing algorithm for depth $3$ ($\Sigma \Pi
\Sigma$) circuits with constant top fan-in. As such, no other general
polynomial time result is known for depth $3$ circuits. A
justification behind the hardness of PIT even for small depth circuits
was provided recently by Agrawal and Vinay \cite{AV08}. They showed
that a deterministic black box identity test for depth $4$ ($\Sigma
\Pi \Sigma \Pi$) circuits would imply a quasi-polynomial time
deterministic PIT algorithm for any circuit computing a polynomial of
\emph{low} degree\footnote{A polynomial is said to have low degree if
  its degree is less than the size of the circuit}. \\

Thus we see that the non-trivial case for identity testing starts with
depth $3$ circuits; whereas circuits of depth $4$ are \emph{almost}
the general case. It is therefore natural to ask as to what is the
complexity of the PIT problem for depth $2$ ($\Pi \Sigma$) circuits if
we allow the \emph{constants} of the circuit to come from an \emph{algebra} \footnote{In this paper we always mean a finite dimensional associative algebra with unity.}
$\mathcal{R}$, which is not a field, and has dimension over $\mathbb{F}$, $\dim_{\F}{(\mathcal{R})} > 1$.  We can make the reasonable
assumption that the algebra $\mathcal{R}$ is given in \emph{basis
  form} i.e. we know an $\mathbb{F}$-basis $\{e_1, \ldots, e_k\}$ of
$\mathcal{R}$ and we also know how $e_i e_j$ can be expressed in terms
of the basis elements, for all $i$ and $j$.  Therefore, the problem at
hand is the following,
\begin{problem}
Given an expression,
\begin{equation*}
P = \prod_{i=1}^{d}{\inparen{A_{i 0} + A_{i 1} x_1 + \ldots + A_{i n}  x_n}}
\end{equation*}
where $A_{i j} \in \mathcal{R}$, an algebra over $\mathbb{F}$ given in
basis form, check if $P$ is zero.
\end{problem}

How hard is the above problem? At first sight, this problem might look
deceptively simple. For instance, if $\mathcal{R}$ is field or even a division
algebra (say, the real quaternion algebra) then it is trivial to check
if $P=0$ using polynomial number of $\mathbb{F}$-operations. However, in general this is far from what might be the
case. \\

Since elements of a finite dimensional algebra, given in basis form,
can be expressed as matrices over $\F$ we can equivalently write the
above problem as,
\begin{problem} \label{prob:mainprob}
Given an expression,
\begin{equation}{\label{eqn:pi_sigma}}
P = \prod_{i=1}^{d}{\inparen{A_{i 0} + A_{i 1} x_1 + \ldots + A_{i n}  x_n}}
\end{equation}
where $A_{i j} \in \M_k(\mathbb{F})$, the algebra of $k \times k$
matrices over $\F$, check if $P$ is zero using $\poly(knd)$ number of $\mathbb{F}$-operations.
\end{problem}

\noindent In order to avoid confusion we would use the following
convention in this paper: \\

\noindent \textbf{Convention} - Whenever we say `arithmetic circuit'
or `arithmetic formula' without any extra qualification, we would mean
a circuit or a formula over a field.  Otherwise, we would explicitly
mention `arithmetic circuit (or formula) over \emph{some} algebra' to
mean that the constants of the circuit are taken from `that' algebra.

\subsection{The depth $2$ model of computation}

A depth $2$ circuit $C$ over matrices, as in Equation
\ref{eqn:pi_sigma}, naturally defines a computational
model. Assuming $\mathcal{R} = \M_k(\mathbb{F})$, for some $k$, 
a polynomial $P \in \mathcal{R}[x_1, \ldots,x_n]$ outputted by $C$ can be viewed as a $k \times k$
matrix of polynomials in $\F[x_1, \ldots, x_n]$. We say
that a polynomial $f \in \F[x_1, \ldots, x_n]$ is \emph{computed} by
$C$ if one of the $k^2$ polynomials in $P$ is $f$.
Sometimes we would abuse terminology a bit and say that $P$
\emph{computes} $f$ to mean the same.  

In the following discussion, we
would denote the algebra of upper-triangular $k \times k$ matrices by
$\mathsf{U}_k(\mathbb{F})$. The algebra $\mathsf{U}_2(\mathbb{F})$ is the 
\emph{smallest} noncommutative algebra with unity over $\mathbb{F}$, in the sense that 
$\dim_{\mathbb{F}}{\mathsf{U}_2(\mathbb{F})} = 3$ whereas any algebra with unity of dimension less than
$3$ is commutative. We show in this paper that already $\mathsf{U}_2(\mathbb{F})$ captures an open case of identity testing. \\

Ben-Or and Cleve \cite{BC88} showed that a polynomial computed by an
arithmetic formula $E$ of depth $d$, and fan-in (of every gate)
bounded by $2$, can also be computed by a straight-line program of
length at most $4^d$ using only $3$ registers. The following fact can
be readily derived from their result (see Theorem \ref{thm:BC}): From
an arithmetic formula $E$ of depth $d$ and fan-in bounded by $2$, we
can efficiently compute the expression,
\begin{equation*} 
P = \prod_{i=1}^{m}{\inparen{A_{i 0} + A_{i 1} x_1 + \ldots + A_{i n}  x_n}}
\end{equation*}
where $m \leq 4^d$ and $A_{i j} \in \M_3(\mathbb{F})$ such that $P$
\emph{computes} the polynomial that $E$ does. Thus solving
Problem~\ref{prob:mainprob} in polynomial time even for $3 \times 3$ matrices
yields a polynomial time algorithm for PIT of constant depth circuits, in 
particular depth $4$ circuits. There is another
way of arguing that the choice of $\mathcal{R}$ as $\M_3(\mathbb{F})$
is \emph{almost} the general case.

For an arithmetic circuit of size $s$, computing a low degree
polynomial, we can use the depth-reduction result by 
Allender, Jiao, Mahajan and Vinay \cite{AJMV98} (see also \cite{VSBR83})
to get an equivalent bounded fan-in formula of size $s^{O(\log s)}$
and depth $O(\log^2 s)$. From this formula we can obtain a depth $2$
circuit over $\M_3(\mathbb{F})$ of size $4^{O(\log^2s)} = s^{O(\log
  s)}$ (using Ben-Or and Cleve's result) that computes the same
polynomial as the formula. Thus, PIT for depth $2$ circuits over $3
\times 3$ matrices is \emph{almost} the general case since a
derandomization yields a quasi-polynomial time PIT algorithm for any
circuit computing a low degree polynomial. This means, in essence a
depth $2$ circuit over $\M_3(\mathbb{F})$ plays the role of a depth
$4$ circuit over $\F$ (using Agrawal and Vinay's result).

What is natural to ask is how the complexity of PIT for depth $2$
circuits over $\M_2(\mathbb{F})$ relates to PIT for arithmetic
circuits. In this paper, we provide an answer to this. We show a
surprising connection between PIT of depth $2$ circuits over
$\mathsf{U}_2(\mathbb{F})$ and PIT of depth $3$ circuits. The reason
this is a bit surprising is because we also show that, a depth $2$
circuit over $\mathsf{U}_2(\mathbb{F})$ is not even powerful enough to
compute a simple polynomial like, $x_1x_2 + x_3x_4 + x_5x_6$!

\subsubsection*{Known related models}

Polynomial identity testing and circuit lower bounds have been studied 
for different algebraic models. Nisan \cite{Nis91} showed an exponential 
lower bound on the size of any arithmetic formula computing the determinant 
of a matrix in the non-commutative \emph{free algebra} model. The result was later generalized 
by Chien and Sinclair \cite{CS04} to a large class of non-commutative algebras
satisfying polynomial identities, called PI-algebras. Identity testing has 
also been studied for the non-commutative model by Raz and Shpilka 
\cite{RS04}, Bogdanov and Wee \cite{BW05}, and Arvind, Mukhopadhyay and
Srinivasan \cite{AMS08}. But unlike this model where the variables do not 
commute, in our setting the variables always commute but the constants 
are taken from an algebra $\mathcal{R}$. The motivation for studying this later model 
is not only because it is a natural generalization of commutative circuits 
over fields but also because it gives a different perspective to the 
complexity of the classical PIT problem in terms of the dimension of the 
underlying algebra $\mathcal{R}$. It seems to `pack' the combinatorial
nature of the circuit into a larger base algebra and hence opens up
the possibility of using algebra structure results. The simplest nontrivial 
circuit in this model is a depth $2$ circuit over the smallest non-commutative 
algebra which is $\mathcal{R} = \mathsf{U}_2(\mathbb{F})$.

\subsection{Our Results}
The results we give are of two types. Some are related to identity
testing while the rest are related to the weakness of the depth $2$
computational model over $\mathsf{U}_2(\mathbb{F})$ and
$\M_2(\mathbb{F})$.

\subsubsection*{{Identity testing}}
We fill in the missing information about the complexity of identity
testing for depth $2$ circuits over $2 \times 2$ matrices by showing
the following result.

\begin{theorem}\label{thm:main-thm}
Identity testing for depth $2$ ($\Pi \Sigma$) circuits over
$\mathsf{U}_2(\mathbb{F})$ is polynomial time equivalent to identity
testing for depth $3$ ($\Sigma \Pi \Sigma$) circuits.
\end{theorem}

\noindent
The above result has an interesting consequence on identity testing
for Algebraic Branching Program (ABP) \cite{Nis91} (see Definition
\ref{defn:ABP}). It is known that identity testing for non-commutative
ABP can be done in deterministic polynomial time (a result due to Raz
and Shpilka \cite{RS04}). But no interesting result is known for
identity testing of even width-2 commutative ABP's. The following
result explains why this is the case.

\begin{corollary}\label{cor:d3w2abp}
Identity testing of depth $3$ circuits reduces to identity testing of
width-$2$ planar ABPs.
\end{corollary}

\noindent Further, we give a deterministic polynomial time identity
testing algorithm for depth $2$ circuits over any constant dimensional
commutative algebra given in basis form. Recall that an algebra
$\mathcal{R}$ is given in basis form if we know an $\mathbb{F}$-basis
$\{e_1, \ldots, e_k\}$ of $\mathcal{R}$ and we also know how $e_i e_j$
can be expressed in terms of the basis elements, for all $i$ and
$j$. Our result can be formally stated as follows.

\begin{theorem} \label{thm:commthm}
Given an expression,
\begin{equation*}
P = \prod_{i=1}^{d}{\inparen{A_{i 0} + A_{i 1} x_1 + \ldots + A_{i n}  x_n}}
\end{equation*}
where $A_{i j} \in \mathcal{R}$, a commutative algebra of constant
dimension over $\mathbb{F}$ that is given in basis form, there is a
deterministic polynomial time algorithm to test if $P$ is zero.
\end{theorem}

\noindent In a way, this result establishes the fact that the power of
depth $2$ $(\Pi \Sigma)$ circuits over constant dimensional algebras 
is primarily derived from the non-commutative nature of the algebra. 
However, things can be very different for commutative algebras of polynomial 
dimension over $\mathbb{F}$. 

\begin{theorem} \label{thm:commthm2}
Identity testing of a depth $3$ $(\Sigma \Pi \Sigma)$ circuit $C$ reduces to 
identity testing of a depth $2$ $(\Pi \Sigma)$ circuit over a commutative algebra 
of dimension polynomial in the size of $C$.
\end{theorem} 

It would be apparent from the proof of Theorem \ref{thm:main-thm} that
our argument is simple in nature. Perhaps the reason why such a
connection was overlooked before is that, unlike a depth $2$ circuit
over $\M_3(\mathbb{F})$, we do not always have the privilege of
\emph{exactly} computing a polynomial over $\mathbb{F}$ using a depth
$2$ circuit over $\mathsf{U}_2(\mathbb{F})$. Showing this weakness of
the latter computational model constitutes the other part of our
results.

\subsubsection*{Weakness of the depth $2$ model over $\mathsf{U}_2(\mathbb{F})$ and $\M_2(\mathbb{F})$}

Although Theorem~\ref{thm:main-thm} shows an equivalence of depth $3$
circuits and depth $2$ circuits over $\mathsf{U}_2(\mathbb{F})$ with
respect to PIT, the computational powers of these two models are very
different. The following result shows that a depth $2$ circuit over
$\mathsf{U}_2(\mathbb{F})$ is computationally strictly weaker than
depth $3$ circuits.

\begin{theorem}\label{thm:d2m<d3}
Let $f \in F[x_1, \ldots, x_n]$ be a polynomial such that there are no
two linear functions $l_1$ and $l_2$ (with $1 \not \in (l_1,l_2)$, the
ideal generated by $l_1$ and $l_2$) which make $f \bmod
\inparen{l_1,l_2}$ also a linear function. Then $f$ is not computable
by a depth $2$ circuit over $\mathsf{U}_2(\mathbb{F})$.
\end{theorem}

It can be shown that even a simple polynomial like $x_1x_2 + x_3x_4 +
x_5x_6$ satisfies the condition stated in the above theorem (see Corollary \ref{cor:explicitfn}), and hence
it is not computable by any depth $2$ circuit over
$\mathsf{U}_2(\mathbb{F})$, no matter how large! This contrast makes
Theorem~\ref{thm:main-thm} surprising as it establishes an equivalence
of identity testing in two models of different
computational strengths. \\

At this point, it is natural to investigate the computational power of
depth $2$ circuits if we graduate from $\mathsf{U}_2(\mathbb{F})$ to
$\M_2(\mathbb{F})$. The following result hints that even such a model
is severely restrictive in nature.

A depth $2$ circuit over $\M_2(\mathbb{F})$ gives as output a
polynomial $P = \prod_{i=1}^{d}{\inparen{A_{i0} + A_{i1}x_1 + \ldots +
    A_{i n}x_n}}$, with $A_{ij} \in \M_2(\mathbb{F})$. Let $P_\ell$
denote the partial product $P_\ell =
\prod_{i=\ell}^{d}{\inparen{A_{i0} + A_{i1}x_1 + \ldots + A_{i
      n}x_n}}$ where $\ell \leq d$.

\begin{definition}\label{defn:degres}
A polynomial $f \in \mathbb{F}[x_1, \ldots, x_n]$ is computed by a
depth $2$ circuit over $\M_2(\mathbb{F})$ \emph{under a degree
  restriction} of $m$ if the degree of each of the partial products
$P_\ell$ is bounded by $m$.
\end{definition}

\begin{theorem}\label{thm:degres}
There exists a class of polynomials of degree $n$ that cannot be
computed by a depth $2$ circuit over $\M_2(\mathbb{F})$, under a
degree restriction of $n$.
\end{theorem}

The motivation for imposing a condition like degree restriction comes
very naturally from depth $2$ circuits over $\M_3(\mathbb{F})$. Given
a polynomial $f = \sum_i{m_i}$, where $m_i$'s are the monomials of
$f$, it is easy to construct a depth $2$ circuit over
$\M_3(\mathbb{F})$ that literally forms these monomials and adds then
one by one. This computation is degree restricted, if we extend our
definition of degree restriction to $\M_3(\mathbb{F})$. However, the
above theorem suggests that no such scheme to compute $f$ would
succeed over $\M_2(\mathbb{F})$. \\

\noindent \textbf{Remark}- By transferring the complexity of an
arithmetic circuit from its depth to the dimension of the underlying
algebras while fixing the depth to $2$, our results provide some
evidence that identity testing for depth $3$ circuits appears to be
\emph{mathematically} more tractable than depth $4$ circuits. Besides,
it might be possible to exploit the properties of these underlying
algebras to say something useful about identity testing. A glimpse of
this indeed appears in our identity testing algorithm over commutative
algebras of constant dimension over $F$.

\subsection{Organization}
The results on identity testing are given in sections \ref{sec:id} and
\ref{sec:commid}, while those on the weakness of the depth $2$ model
are given in section \ref{sec:weakness}.  In section \ref{sec:id} we
prove the equivalence of identity testing between depth $3$ circuits
and depth $2$ circuits over $\mathsf{U}_2(\mathbb{F})$ (Theorem
\ref{thm:main-thm}), and show how it connects to width-$2$ ABPs
(Corollary \ref{cor:d3w2abp}). The deterministic polynomial time
identity testing algorithm over commutative algebra of constant dimension 
is presented in section \ref{sec:commid} (Theorem \ref{thm:commthm}). In the same 
section it is also shown that commutative algebras of polynomial dimensions are 
powerful enough to capture PIT of depth $3$ $(\Sigma \Pi \Sigma)$ circuits (Theorem \ref{thm:commthm2}).
Finally, in section \ref{sec:weakness} we show the weakness of the depth $2$ model
over $\mathsf{U}_2(\mathbb{F})$ and $\M_2(\mathbb{F})$ (Theorem
\ref{thm:d2m<d3} and \ref{thm:degres}).

\section{Identity testing over $\M_2(\mathbb{F})$} \label{sec:id}

In this section, we show that PIT of depth $2$ circuits over
$\M_2(\mathbb{F})$ is at least as hard as PIT of depth $3$ circuits,
and this further implies that PIT of a width-$2$ commutative ABP is
also at least as hard as PIT of depth $3$ circuits.

\subsection{Equivalence with depth $3$ identity testing} 
We will now prove Theorem~\ref{thm:main-thm}. Given a depth $3$
circuit we can assume, without loss of generality, that the fan-in of
the multiplication gates are the same. This multiplicative fan-in will
be referred to as the \emph{degree} of the depth $3$ circuit. The
following lemma is the crux of our argument. For convenience, we will
call a matrix with linear functions as entries, a \emph{linear}
matrix.

\begin{lemma}\label{lem:main-lem}
 Let $f \in \mathbb{F}[x_1, \ldots,x_n]$ be a polynomial computed by a
 depth $3$ circuit $C$ of degree $d$ and top level fan-in $s$. Given
 circuit $C$, it is possible to construct in polynomial time a depth
 $2$ circuit over $\mathsf{U}_2(\mathbb{F})$ of size $O((d+n)s^2)$
 that computes a polynomial $p = L \cdot f$, where $L$ is a product of
 non-zero linear functions.
\end{lemma}

\begin{proof}
  A depth $2$ circuit over $\mathsf{U}_2(\mathbb{F})$ is simply a
  product sequence of $2 \times 2$ upper-triangular linear
  matrices. We now show that there exists such a sequence of length
  $O((d+n)s^2)$ such that the product $2 \times 2$ matrix has $L \cdot
  f$ as one of its entries.

  Since $f$ is computed by a depth $3$ circuit, we can write $f =
  \sum_{i=1}^sP_i$, where each summand $P_i = \prod_{j}l_{ij}$ is a
  product of linear functions.  Observe that we can compute a single
  $P_i$ using a product sequence of length $d$ as:
  \begin{equation} \label{eqn:indbasis}
  \mat{l_{i1} & \\ & 1} \mat{l_{i2} & \\ & 1} \cdots \mat{l_{i(d-1)} &
    \\ & 1} \mat{1 & l_{id}\\ & 1} = \mat{L' & P_i\\ & 1}
  \end{equation}
  where $L' = l_{i1}\cdots l_{i(d-1)}$. 
  
  \noindent
  Each matrix of the form $\mat{1 & l\\ & 1}$, where $l = a_0 + \sum
  a_ix_i$, can be further expanded as,
  $$
  \mat{1 & a_0\\&1}\mat{1 & a_1x_1\\&1}\cdots
  \mat{1 & a_nx_n\\&1} = \mat{1 & l\\ & 1}
  $$ These will be the only type of non-diagonal matrices that would
  appear in the sequence. \\

  The proof will proceed by induction where Equation
  \ref{eqn:indbasis} serves as the induction basis.  A generic
  intermediate matrix would look like $\mat{L_1 & L_2 g\\ & L_3}$
  where each $L_i$ is a product of non-zero linear functions and $g$
  is a partial summand of $P_i$'s. We shall inductively double the
  number of summands in $g$ at each step.

  At the $i$-th iteration let us assume that we have the matrices
  $\mat{L_1 & L_2 g\\ & L_3}$ and $\mat{M_1 & M_2 h\\ & M_3}$, each
  computed by a sequence of $n_i$ linear matrices. We now want a
  sequence that computes a polynomial of the form $L \cdot
  (g+h)$. Consider the following sequence,
  \begin{equation} \label{eqn:indhyp}
  \mat{{L}_1 & {L}_2 g\\ & {L}_3}\mat{{A}& \\ & {B}}\mat{{M}_1 & {M}_2
    h\\ & {M}_3} = \mat{{A}{L}_1 {M}_1 & {A}{L}_1 {M}_2h + {B}{L}_2
    {M}_3g\\ & {B}{L}_3 {M}_3}
  \end{equation}
  where ${A}$, ${B}$ are products of linear functions. By setting ${A}
  = {L}_2 {M}_3$ and ${B} = {L}_1{M}_2$ we get the desired sequence,
  $$ \mat{{L}_1 & {L}_2 g\\ & {L}_3}\mat{{A}& \\ & {B}}\mat{{M}_1 &
    {M}_2 h\\ & {M}_3} = \mat{{L}_1 {L}_2 {M}_1 {M}_3 & {L}_1 {L}_2
    {M}_2 {M}_3 (g+h) \\ & {L}_1 {L}_3 {M}_2 {M}_3}
  $$ This way, we have doubled the number of summands in $g + h$. The
  length of the sequence computing ${L}_2g$ and ${M}_2h$ is $n_i$,
  hence each ${L}_i$ and ${M}_i$ is a product of $n_i$ many linear
  functions. Therefore, both ${A}$ and ${B}$ are products of at most
  $2n_i$ linear functions and the matrix $\mat{{A} & \\ & {B}}$ can be
  written as a product of at most $2n_i$ diagonal linear matrices. The
  total length of the sequence given in Equation~\ref{eqn:indhyp} is
  hence bounded by $4n_i$.

  The number of summands in $f$ is $s$ and the above process needs to
  be repeated at most $\log{s} + 1$ times. The final sequence length
  is hence bounded by $(d+n)\cdot 4^{\log{s}} = (d+n)s^2$.
\end{proof}

\medskip
\begin{proof}[Proof of Theorem~\ref{thm:main-thm}]
It follows from Lemma~\ref{lem:main-lem} that, given a depth $3$
circuit $C$ computing $f$ we can efficiently construct a depth $2$
circuit over $\mathsf{U}_2(\mathbb{F})$ that outputs a matrix,
$\mat{{L}_1 & L \cdot f\\ & {L}_2}$, where $L$ is a product of
non-zero linear functions. Multiplying this matrix by $\mat{1 & 0\\ &
  0}$ to the left and $\mat{0 & 0\\ & 1}$ to the right yields another
depth $2$ circuit $D$ that outputs $\mat{0 & L \cdot f\\ & 0}$. Thus
$D$ computes an identically zero polynomial over
$\mathsf{U}_2(\mathbb{F})$ if and only if $C$ computes an identically
zero polynomial. This shows that PIT for depth $3$ circuits reduces to
PIT of depth $2$ circuits over $\mathsf{U}_2({\mathbb{F}})$.

The other direction, that is PIT for depth $2$ circuits over
$\mathsf{U}_2(\mathbb{F})$ reduces to PIT for depth $3$ circuits, is
trivial to observe. The diagonal entries of the output $2 \times 2$
matrix is just a product of linear functions whereas the off-diagonal
entry is a sum of at most $d'$ many products of linear functions,
where $d'$ is the multiplicative fan-in of the depth $2$ circuit over
$\mathsf{U}_2(\mathbb{F})$.
\end{proof}

\subsection{Width-$2$ algebraic branching programs}
Algebraic Branching Program (ABP) is a model of computation defined by
Nisan \cite{Nis91}. Formally, an ABP is defined as follows.
\begin{definition}\label{defn:ABP}(Nisan~\cite{Nis91}) An \emph{algebraic branching
  program (ABP)} is a directed acyclic graph with one source and one
  sink. The vertices of this graph are partitioned into levels
  labelled $0$ to $d$, where edges may go from level $i$ to level
  $i+1$. The parameter $d$ is called the degree of the ABP. The source
  is the only vertex at level $0$ and the sink is the only vertex at
  level $d$. Each edge is labelled with a homogeneous linear function
  of $x_1, \ldots, x_n$ (i.e. a function of the form $\sum_i{c_ix_i}$). 
  The \emph{width} of the ABP is the maximum number of
  vertices in any level, and the \emph{size} is the total number of vertices.

  An ABP computes a function in the obvious way; sum over all paths
  from source to sink, the product of all linear functions by which
  the edges of the path are labelled. 
\end{definition}

An ABP is said to be \emph{planar} if the underlying graph is
planar. \\

\noindent The following argument shows how Corollary \ref{cor:d3w2abp}
follows easily from Theorem \ref{thm:main-thm}.
\noindent
\begin{proof}[Proof of Corollary~\ref{cor:d3w2abp}. ]
\noindent Theorem \ref{thm:main-thm} constructs a depth $2$ circuit
$D$ that computes $P = \prod_{j}(A_{j 0} + A_{j 1}x_1 + \ldots + A_{j
  n}x_n)$, where each $A_{j i} \in \mathsf{U}_2(\mathbb{F})$. We can
make $D$ homogeneous by introducing an extra variable $z$, such that
$P = \prod_{j}(A_{j 0}z + A_{j 1}x_1 + \ldots + A_{j n}x_n)$. This
means, the product sequence considered in Lemma \ref{lem:main-lem}, is
such that all the linear matrices have homogeneous linear functions as
entries and the only non-diagonal linear matrices are of the form
$\mat{z&cx_i\\&z}$. It is now straightforward to construct a width-$2$
ABP by making the $j^{th}$ linear matrix in the sequence act as the
adjacency matrix between level $j$ and $j+1$ of the ABP. The ABP
constructed is planar since it has layers only of the following two
kinds:

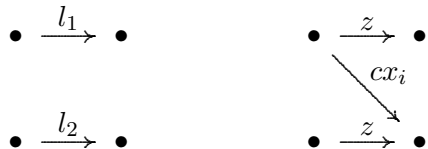
\begin{eqnarray*}
{\begindc{\commdiag}[40]
\obj(0,0){$\bullet$}
\obj(0,1){$\bullet$}
\obj(1,0){$\bullet$}
\obj(1,1){$\bullet$}
\mor(0,0)(1,0){$l_2$}
\mor(0,1)(1,1){$l_1$}
\enddc}
&\qquad\qquad &
{\begindc{\commdiag}[40]
\obj(0,0){$\bullet$}
\obj(0,1){$\bullet$}
\obj(1,0){$\bullet$}
\obj(1,1){$\bullet$}
\mor(0,0)(1,0){$z$}
\mor(0,1)(1,1){$z$}
\mor(0,1)(1,0){$cx_i$}
\enddc}
\end{eqnarray*}
where $l_1$. $l_2$ are homogeneous linear functions.  
\end{proof}

As a matter of fact, the above argument actually shows that PIT of
depth $2$ circuits over $\M_2(\mathbb{F})$ reduces to PIT of width-2
ABPs.

\section{Identity testing over commutative algebras} \label{sec:commid}

\noindent We would now prove Theorem~\ref{thm:commthm}. The main idea
behind this proof is a structure theorem for finite dimensional
commutative algebras over a field. To state the theorem we need the
following definition.
\begin{definition}
A ring $\mathcal{R}$ is \emph{local} if it has a unique maximal ideal.
\end{definition}
An element $u$ in a ring $\mathcal{R}$ is said to be a \emph{unit} if
there exist an element $u'$ such that $uu' = 1$, where $1$ is the
identity element of $\mathcal{R}$. An element $m \in \mathcal{R}$ is
\emph{nilpotent} if there exist a positive integer $n$ with $m^n =
0$. In a local ring the unique maximal ideal consists of all non-units
in $\mathcal{R}$. \\

The following theorem shows how a commutative algebra
decomposes into local sub-algebras. The theorem is quite well known in
the theory of commutative algebras. But since we need an effective
version of this theorem, we present the proof here for the sake of
completion and clarity.
\begin{theorem} \label{thm:localthm}
A finite dimensional commutative algebra $\mathcal{R}$ over
$\mathbb{F}$ is isomorphic to a direct product of
local rings i.e.
\begin{equation*}
\mathcal{R} \cong \mathcal{R}_1 \oplus \ldots \oplus \mathcal{R}_{\ell}
\end{equation*}
where each $\mathcal{R}_i$ is a local ring contained in $\mathcal{R}$
and any non-unit in $\mathcal{R}_i$ is nilpotent.
\end{theorem}
\begin{proof}
If all non-units in $\mathcal{R}$ are nilpotents then $\mathcal{R}$ is
a local ring and the set of nilpotents forms the unique maximal
ideal. Therefore, suppose that there is a non-nilpotent zero-divisor
$z$ in $\mathcal{R}$. (Any non-unit $z$ in a finite dimensional
algebra is a zero-divisor i.e. $\exists y \in \mathcal{R}$ and $y \neq
0$ such that $y z = 0$.) We would argue that using $z$ we can find an
\emph{idempotent} $v \not \in \{ 0, 1 \}$ in $\mathcal{R}$ i.e. $v^2 =
v$. 

Assume that we do have a non-trivial idempotent $v \in
\mathcal{R}$. Let $\mathcal{R}v$ be the sub-algebra of $\mathcal{R}$
generated by multiplying elements of $\mathcal{R}$ with $v$. Since any
$a = av + a(1 - v)$ and $\mathcal{R}v \cap \mathcal{R}(1 - v) =
\inbrace{0}$, we get $\mathcal{R} \cong \mathcal{R}v \oplus
\mathcal{R}(1 - v)$ as a non-trivial decomposition of
$\mathcal{R}$. (Note that $\mathcal{R}$ is a direct sum of the two
sub-algebras because for any $a \in \mathcal{R}v$ and $b \in
\mathcal{R}(1-v)$, $a \cdot b = 0$. This is the place where we use
commutativity of $\mathcal{R}$.) By repeating the splitting process on
the sub-algebras we can eventually prove the theorem. We now show how to find
an idempotent from the zero-divisor $z$.\\

An element $a \in \mathcal{R}$ can be expressed equivalently as a
matrix in $\M_k(\mathbb{F})$, where $k =
\dim_{\mathbb{F}}(\mathcal{R})$, by treating $a$ as the linear
transformation on $\mathcal{R}$ that takes $b \in \mathcal{R}$ to $a
\cdot b$. Therefore, $z$ is a zero-divisor if and only if $z$ as a
matrix is singular. Consider the Jordan normal form of $z$. Since it
is merely a change of basis we would assume, without loss of
generality, that $z$ is already in Jordan normal form. (We won't
compute the Jordan normal form in our algorithm, it is used only for
the sake of argument.) Let,
\begin{equation*}
z = \insquar{\begin{array}{cc}
A & 0 \\
0 & N
\end{array} 
}
\end{equation*}
where $A, N$ are block diagonal matrices and $A$ is non-singular and
$N$ is nilpotent. Therefore there exits a positive integer $t < k$
such that,
\begin{equation*}
w = z^t = \insquar{\begin{array}{cc}
B & 0 \\
0 & 0
\end{array} 
}
\end{equation*}
where $B = A^t$ is non-singular. The claim is, there is an identity element 
in the sub-algebra $\mathcal{R}w$ which can be taken to be the idempotent that splits $\mathcal{R}$. To see this first observe that the
minimum polynomial of $w$ over $\F$ is $m(x) = x\cdot m'(x)$,
where $m'(x)$ is the minimum polynomial of $B$. Also if $m(x) =
{\sum_{i=1}^k\alpha_i x^i}$ then $\alpha_1\neq 0$ as it is the constant term of $m'(x)$ and $B$ is non-singular. Therefore, there
exists an $a \in \mathcal{R}$ such that $w \cdot (aw - 1) = 0$. We can
take $v = aw$ as the identity element in the sub-algebra
$\mathcal{R} w$. This $v \not \in \inbrace{0,1}$ is the 
required idempotent in $\mathcal{R}$.
\end{proof}

\noindent We are now ready to prove Theorem~\ref{thm:commthm}.

\medskip
\noindent
{\bf Theorem~\ref{thm:commthm}} (restated.) {\sl
Given an expression,
\begin{equation*}
P = \prod_{i=1}^{d}{\inparen{A_{i 0} + A_{i 1} x_1 + \ldots + A_{i n}  x_n}}
\end{equation*}
where $A_{i j} \in \mathcal{R}$, a commutative algebra of constant
dimension over $\mathbb{F}$ that is given in basis form, there is a
deterministic polynomial time algorithm to test if $P$ is zero.  }
\begin{proof}
Suppose, the elements $e_1, \ldots, e_k$ form a basis of $\mathcal{R}$
over $\mathbb{F}$. Since any element in $\mathcal{R}$ can be
equivalently expressed as a $k \times k$ matrix over $\mathbb{F}$ (by
treating it as a linear transformation), we will assume that $A_{i j}
\in \M_k(\mathbb{F})$, for all $i$ and $j$. Further, since
$\mathcal{R}$ is given in basis form, we can find these matrix
representations of $A_{i j}$'s efficiently. 

If every $A_{i j}$ is non-singular, then surely $P \neq 0$. (This can be argued by fixing an ordering $x_1 \succ x_2 \succ \ldots \succ x_n$ among the variables. The coefficient of the leading monomial of $P$, with respect to this ordering, is a product of invertible matrices and hence $P \neq 0$.) Therefore,
assume that $\exists A_{i j} = z$ such that $z$ is a zero-divisor
i.e. singular. From the proof of Theorem \ref{thm:localthm} it follows
that there exists a $t < k$ such that the sub-algebra $\mathcal{R}w$,
where $w = z^t$, contains an identity element $v$ which is an
idempotent. To find the right $w$ we can simply go through all $1\leq
t < k$. We now argue that for the correct choice of $w$, $v$ can be
found by solving a system of linear equations over $\mathbb{F}$. Let
$b_1, \ldots, b_{k'}$ be a basis of $\mathcal{R}w$, which we can find
easily from the elements $e_1w, \ldots, e_kw$. In order to solve for
$v$ write it as,
\begin{equation*}
v = \nu_1 b_1 + \ldots + \nu_{k'} b_{k'} 
\end{equation*}
where $\nu_j \in \mathbb{F}$ are unknowns.  Since $v$ is an identity in $\mathcal{R}w$ we
have the following equations,
\begin{equation*}
\left( \nu_1 b_1 + \ldots + \nu_{k'} b_{k'} \right) \cdot b_i = b_i
\text{\hspace{0.1in} for $1 \leq i \leq k'$.}
\end{equation*}
Expressing each $b_i$ in terms of $e_1, \ldots, e_k$, we get a set of
linear equations in $\nu_j$'s. Thus for the right choice of $w$
(i.e. for the right choice of $t$) there is a solution for $v$.  On
the other hand, a solution for $v$ for any $w$ gives us an idempotent,
which is all that we need.

Since $\mathcal{R} \cong \mathcal{R}v \oplus \mathcal{R}(1 - v)$
we can now split the identity testing problem into two similar problems,
i.e. $P$ is zero if and only if,
\begin{eqnarray*}
Pv &=& \prod_{i=1}^{d}{\left(A_{i 0}v + A_{i 1}v \cdot x_1 + \ldots +
  A_{i n}v \cdot x_n \right)} \text{\hspace{0.1in} and} \\ P(1-v) &=&
\prod_{i=1}^{d}{\left(A_{i 0}(1 - v) + A_{i 1}(1 - v) \cdot x_1 +
  \ldots + A_{i n}(1- v) \cdot x_n \right)}
\end{eqnarray*}
are both zero. What we just did with $P \in \mathcal{R}$ we can repeat
for $Pv \in \mathcal{R}v$ and $P(1-v) \in \mathcal{R}(1-v)$. By
decomposing the algebra each time an $A_{i j}$ is a non-nilpotent
zero-divisor, we have reduced the problem to the following easier
problem of checking if
\begin{equation*}
P = \prod_{i=1}^{d}{\left(A_{i 0} + A_{i 1} x_1 + \ldots + A_{i n}  x_n \right)}
\end{equation*}
is zero, where the coefficients $A_{i j}$'s are either nilpotent or
invertible matrices. 

Let $T_i = \left(A_{i 0} + A_{i 1} x_1 + \ldots + A_{i n}  x_n \right)$ 
be a term such that the coefficient of $x_j$ in $T_i$, i.e. $A_{ij}$ is invertible. And suppose 
$Q$ be the product of all terms other than $T_i$. Then $P = T_i \cdot Q$ (since $\mathcal{R}$ is commutative). 
Fix an ordering among the variables so that $x_j$ gets the highest priority. The leading coefficient of 
$P$, under this ordering, is $A_{ij}$ times the leading coefficient of $Q$. Since $A_{ij}$ is invertible this implies 
that $P = 0$ if and only if $Q = 0$. (If $A_{i 0}$ is invertible, we can arrive at the same conclusion by arguing with the coefficients of the least monomials of $P$ and $Q$ under some ordering.) In other words, $P = 0$ if
and only if the product of all those terms for which all the coefficients are nilpotent matrices is
zero. But this is easy to
check since the dimension of the algebra, $k$ is a constant. (In fact,
this is the only step where we use that $k$ is a constant.)  If number
of such terms is greater than $k$ then $P$ is automatically zero (this follows
easily from the fact that the commuting nilpotent matrices can be simultaneously
triangularized with zeroes in the diagonal). Otherwise, simply multiply those terms and check if
it is zero. This takes $O(n^k)$ operations over $\mathbb{F}$.
\end{proof}

\noindent It is clear from the above discussion that identity testing of depth $2$ $(\Pi \Sigma)$
circuits over commutative algebras reduces in polynomial time to that over local rings. As long 
as the dimensions of these local rings are constant we are through. But what happens for nonconstant  
dimensions? The following result 
justifies the hardness of this problem.

\medskip
\noindent
{\bf Theorem~\ref{thm:commthm2}} (restated.) {\sl Given a depth $3$ $(\Sigma \Pi \Sigma)$ 
circuit $C$ of degree $d$ and top level fan-in $s$, it is possible to construct in polynomial time 
a depth $2$ $(\Pi \Sigma)$ 
circuit $\tilde{C}$ over a local ring of dimension $s(d - 1) + 2$ over $\mathbb{F}$ such 
that $\tilde{C}$ computes a zero polynomial if and only if $C$ does so.}
\begin{proof}
The proof is relatively straightforward. Consider a depth $3$ $(\Sigma \Pi \Sigma)$ circuit computing a 
polynomial $f = \sum_{i = 1}^{s}{\prod_{j=1}^{d}{l_{i j}}}$, where $l_{i j}$'s are linear functions. Consider 
the ring $\mathcal{R} = \mathbb{F}[y_1, \ldots, y_s] / \mathcal{I}$, where $\mathcal{I}$ is an ideal 
generated by the elements $\{ y_i y_j \}_{1 \leq i < j \leq s}$ and $\{ y_1^d - y_i^d \}_{1 < i \leq s}$. Observe 
that $\mathcal{R}$ is a local ring, as $y_i^{d+1} = 0$ for all $1 \leq i \leq s$. Also the elements 
$\{1, y_1, \ldots, y_1^d, y_2, \ldots, y_2^{d-1}, \ldots, y_s, \ldots, y_s^{d-1}\}$ form an $\mathbb{F}$-basis of 
$\mathcal{R}$. Now notice that the polynomial,
\begin{eqnarray*}
P &=& \prod_{j=1}^{d}{\left( l_{j 1} y_1 + \ldots + l_{j s} y_s \right)} \\
&=& f \cdot y_1^d
\end{eqnarray*}
is zero if and only if $f$ is zero. Polynomial $P$ can indeed be computed by a depth $2$ ($\Pi \Sigma$) circuit over $\mathcal{R}$.
\end{proof}

\section{Weakness of the depth $2$ model}\label{sec:weakness}

In Lemma \ref{lem:main-lem}, we saw that the depth $2$ circuit over $\mathsf{U}_2(\mathbb{F})$ 
computes $L \cdot f$ instead of $f$. Is it 
possible to drop the factor $L$ and simply compute $f$? In this section, we show that in \emph{many} cases it is impossible 
to find a depth $2$ circuit over $\mathsf{U}_2(\mathbb{F})$ that computes $f$.

\subsection{Depth $2$ model over $\mathsf{U}_2(\mathbb{F})$}

We will now prove Theorem~\ref{thm:d2m<d3}. In the following discussion we use the notation $(l_1,l_2)$ 
to mean the ideal generated by two linear functions $l_1$ and $l_2$. Further, we say that $l_1$ is \emph{independent} of $l_2$
if $1 \not \in (l_1,l_2)$. \\

\noindent
{\bf Theorem~\ref{thm:d2m<d3}} (restated.) {\sl
Let $f \in F[x_1, \ldots, x_n]$ be a polynomial such that there are no two
linear functions $l_1$ and $l_2$ (with $1 \not \in (l_1,l_2)$) which make $f \bmod \inparen{l_1,l_2}$ also a linear function. Then $f$ is not 
computable by a depth $2$ circuit over $\mathsf{U}_2(\mathbb{F})$.
}
\begin{proof}
  Assume on the contrary that $f$ can be computed by a depth $2$ circuit over 
  $\mathsf{U}_2(\mathbb{F})$. In other words, there is a product sequence $M_1\cdots M_t$ 
  of $2 \times 2$ upper-triangular linear matrices such that $f$ appears as the top-right entry 
  of the final product. Let  $M_i =
  \mat{l_{i1} & l_{i2}\\ & l_{i3}}$, then 
  \begin{equation}\label{eqn:fM_1M_n}
  f = \mat{1 & 0}\mat{l_{11} & l_{12}\\ & l_{13}}\mat{l_{21} &
    l_{22}\\ & l_{23}}\cdots \mat{l_{t1} & l_{t2}\\ &
    l_{t3}}\insquar{\begin{array}{c} 0 \\1\end{array}}
  \end{equation}
  
  \noindent
  {\bf Case $1$: }Not all the $l_{i1}$'s are constants.
  \medskip

  Let $k$ be the least index such that $l_{k1}$ is not a constant and
  $l_{i1} = c_i$ for all $i<k$. To simplify
  Equation~\ref{eqn:fM_1M_n}, let
  \begin{eqnarray*}
    \insquar{\begin{array}{c} B \\ L\end{array}} & = &
    M_{k+1}\cdots M_t\insquar{\begin{array}{c} 0\\1\end{array}}\\
    \mat{d_i&D_i} & = &\mat{1&0} \cdot M_1\cdots M_{i-1}
  \end{eqnarray*}
  Observe that $L$ is just a product of linear functions,
  and for all $1\leq i < k$, we have the following relations.
  \begin{eqnarray*}
    d_{i+1} & = & \prod_{j=1}^ic_j\\
    D_{i+1} & = & d_{i}l_{i2} + l_{i3}D_{i}
  \end{eqnarray*}
  
  \noindent
  Hence, Equation~\ref{eqn:fM_1M_n} simplifies as
  \begin{eqnarray*}
    f & = & \mat{d_k & D_k}\mat{l_{k1} & l_{k2}\\ &
      l_{k3}}\insquar{\begin{array}{c} B\\L\end{array}}\\
    & = & d_kl_{k1}B + \inparen{d_kl_{k2} + l_{k3}D_k}L
  \end{eqnarray*}
  
  Suppose there is some factor $l$ of $L$ with $1 \not \in (l_{k1},l)$. Then $f = 0\bmod{(l_{k1},l)}$, which is not
  possible. Hence, $L$ must be a constant modulo $l_{k1}$. For
  appropriate constants $\alpha,\beta$, we have
  \begin{equation}\label{eqn:indDk}
    f = \alpha l_{k2} + \beta l_{k3}D_k \pmod{l_{k1}}
  \end{equation}

  We argue that the above equation cannot be true by inducting on
  $k$.  If $l_{k3}$ was independent of $l_{k1}$, then $f = \alpha
  l_{k2} \bmod \inparen{l_{k1},l_{k3}}$ which is not
  possible. Therefore, $l_{k3}$ must be a constant modulo $l_{k1}$. We
  then have the following (reusing $\alpha$ and $\beta$ to denote
  appropriate constants):
  \begin{eqnarray*}
    f & = & \alpha l_{k2} + \beta D_k \pmod{l_{k1}}\\
      & = & \alpha l_{k2} + \beta\inparen{d_{k-1}l_{(k-1)2} +
      l_{(k-1)3}D_{k-1}}\pmod{l_{k1}}\\
     \implies f& = & \inparen{\alpha l_{k2}+ \beta d_{k-1}l_{(k-1)2}} + \beta
    l_{(k-1)3}D_{k-1} \pmod{l_{k1}}
  \end{eqnarray*}
  The last equation can be rewritten in the form of
  Equation~\ref{eqn:indDk} with $\beta l_{k3}D_{k}$ replaced by $\beta
  l_{(k-1)3}D_{k-1}$. Notice that the expression $\inparen{\alpha
    l_{k2}+ \beta d_{k-1}l_{(k-1)2}}$ is linear just like $\alpha
  l_{k2}$. Hence by using the argument iteratively we eventually get a contradiction at $D_1$. \\
  
  \noindent
  {\bf Case $2$: }All the $l_{i1}$'s are constants.
  \medskip

  In this case, Equation~\ref{eqn:fM_1M_n} can be rewritten as
  \begin{eqnarray*}
    f & = & \mat{d_t & D_t}\mat{c_t& l_{t2}\\ &
      l_{t3}}\insquar{\begin{array}{c} 0\\1\end{array}}\\
      & = & d_tl_{t2} + l_{t3}D_t
  \end{eqnarray*}
  The last equation is again of the form in
  Equation~\ref{eqn:indDk} (without the mod term) and hence the same
  argument can be repeated here as well to give the desired
  contradiction.
\end{proof}

The following corollary provides some explicit examples of functions
that cannot be computed. 

\begin{corollary} \label{cor:explicitfn}
A depth $2$ circuit over $\mathsf{U}_2(\mathbb{F})$ cannot
compute the polynomial $x_1x_2 + x_3x_4 + x_5x_6$. Other examples include well known functions like
$\mathsf{det}_n$ and $\mathsf{perm}_n$, 
the determinant and permanent polynomials, for $n\geq 3$.
\end{corollary}
\begin{proof}
  It suffices to show that $f = x_1x_2 +x_3x_4+x_5x_6$ satisfy the
  requirement in Theorem~\ref{thm:d2m<d3}. 

  To obtain a contradiction, let us assume that there does 
  exist two linear functions $l_1$ and $l_2$ (with $1 \not \in (l_1,l_2)$) such that
  $f\bmod{(l_1,l_2)}$ is linear. We can evaluate $f\bmod{(l_1,l_2)}$ 
  by substituting a pair of the variables in $f$ by linear
  functions in the rest of the variables (as dictated by the equations $l_1 = l_2 = 0$). By the symmetry of $f$, we can assume that the pair
  is either $\inbrace{x_1,x_2}$ or $\inbrace{x_1,x_3}$.

  If $x_1 = l'_1$ and $x_3 = l'_2$ are the substitutions, then $l'_1x_2 +
  \l'_2x_4$ can never contribute a term to cancel off $x_5x_6$ and
  hence $f \bmod{(l_1,l_2)}$ cannot be linear.

  Otherwise, let $x_1 = l'_1$ and $x_2 = l'_2$ be the substitutions. If
  $f\bmod{(l_1,l_2)} = l'_1l'_2 + x_3x_4 + x_5x_6$ is linear,
  there cannot be a common $x_i$ with non-zero coefficient in both
  $l'_1$ and $l'_2$.  Without loss of generality, assume that
  $l'_1$ involves $x_3$ and $x_5$ and $l'_2$ involves $x_4$ and
  $x_6$. But then the product $l'_1l'_2$ would involve terms like $x_3x_6$ that cannot be
  cancelled, contradicting linearity again.
\end{proof}

\subsection{Depth $2$ model over $\M_2(\mathbb{F})$}

In this section we show that the power of depth $2$ circuits is very restrictive even if we take 
the underlying algebra to be $\M_2(\mathbb{F})$ instead of $\mathsf{U}_2(\mathbb{F})$. 
In the following discussion, we will refer to a homogeneous linear function as a \emph{linear form}.  

\begin{definition}
A polynomial $f$ of degree $n$ is said to be
\emph{$r$-robust} if $f$ does not belong to any ideal generated by $r$
linear forms.
\end{definition}

\noindent For instance, it can be checked that $\mathsf{det}_n$ and
$\mathsf{perm}_n$, the symbolic determinant and permanent of an $n
\times n$ matrix, are $(n-1)$-robust polynomials. For any
polynomial $f$, we will denote the $d^{th}$ homogeneous part of $f$ by
$[f]_d$. And let $(h_1,\cdots, h_k)$ denote the ideal
generated by $h_1,\cdots, h_k$. For the following theorem recall 
the definition of \emph{degree restriction} (Definition \ref{defn:degres}) given in Section \ref{sec:intro}.

\begin{theorem}\label{thm:comp-newer}
A polynomial $f$ of degree $n$, such that $[f]_n$ is $5$-robust,
cannot be computed by a depth $2$ circuit over $\M_2(\mathbb{F})$
under a degree restriction of $n$.
\end{theorem}

\noindent We prove this with the help of the following lemma, which basically applies
Gaussian column operations to simplify matrices.

\begin{lemma}\label{lem:degres}
Let $f_1$ be a polynomial of degree $n$ such that $[f_1]_n$ is
$4$-robust. Suppose there is a linear matrix $M$ and polynomials
$f_2,g_1,g_2$ of degree at most $n$ satisfying
$$ \insquar{\begin{array}{c} f_1\\f_2\end{array}} =
M\insquar{\begin{array}{c} g_1\\g_2\end{array}}
$$ Then, there is an appropriate invertible column operation $A$ such
that 
$$
M\cdot A = \mat{1 & h_2\\c_3 & h_4 + c_4}
$$
where $c_3, c_4$ are constants and $h_2,h_4$ are linear 
forms.
\end{lemma}

\noindent
We will defer the proof of this lemma to the end of this section, and
shall use it to prove Theorem~\ref{thm:comp-newer}. 

\begin{proof}[Proof of Theorem~\ref{thm:comp-newer}]
Assume, on the contrary, that we do have such a sequence of matrices
computing $f$. Since only one entry is of interest to us, we shall
assume that the first matrix is a row vector and the last matrix is a
column vector. Let the sequence of minimum length computing $f$ be the
following:
$$ f = \bar{v}\cdot M_1M_2\cdots M_d\cdot \bar{w}
$$

Using Lemma~\ref{lem:degres} we shall repeatedly transform the above
sequence by replacing $M_iM_{i+1}$ by $(M_iA)(A^{-1}M_{i+1})$ for an
appropriate invertible column transformation $A$. Since $A$ would
consist of just constant entries, $M_iA$ and $A^{-1}M_{i+1}$ continue to
be linear matrices. \\

To begin, let $\bar{v} = [l_1, l_2]$ for two linear functions $l_1$
and $l_2$. And let $\insquar{f_1,f_2}^T = M_1\cdots M_d\bar{w}$. Then
we have, 
$$ \insquar{\begin{array}{c}f\\0\end{array}} = \mat{l_1 & l_2\\0 &
  0}\insquar{\begin{array}{c}f_1\\f_2\end{array}}
$$ Hence, by Lemma~\ref{lem:degres}, we can assume $\bar{v} = [1, h]$
and hence $f = f_1 + hf_2$. By the minimality of the sequence, $h\neq
0$. This forces $f_1$ to be $4$-robust and the degree restriction
makes $[f_2]_n=0$.

Let $[g_1,g_2]^T = M_2\cdots M_d\bar{w}$. The goal is to translate the
properties that $[f_1]_n$ is $4$-robust and $[f_2]_n =0$ to the
polynomials $g_1$ and $g_2$. Translating these properties would show
each $M_i$ is of the form described in Lemma~\ref{lem:degres}. Thus,
inducting on the length of the sequence, we would arrive at the
required contradiction. In general, we have an equation of the form
$$ \insquar{\begin{array}{c} f_1\\f_2\end{array}} =
M_i\insquar{\begin{array}{c} g_1\\g_2\end{array}}
$$ Since $[f_1]_n$ is $4$-robust, using Lemma~\ref{lem:degres} again,
we can assume that
\begin{equation}\label{eqn:degres}
 \insquar{\begin{array}{c} f_1\\f_2\end{array}} = \mat{1 & h_2\\c_3
  & c_4 + h_4}\insquar{\begin{array}{c} g_1\\g_2\end{array}}
\end{equation}
by reusing the variables $g_1$, $g_2$ and others. 
Observe that in the above equation if $h_4 = 0$ then $M_{i-1}M_i$ still continues to be a
linear matrix (since, by induction, $M_{i-1}$ is of the form as dictated by Lemma \ref{lem:degres}) and that would contradict the minimality of the
sequence. Therefore $h_4\neq 0$.\\

\noindent
\emph{Claim: }$c_3 = 0$ (by comparing the $n^{th}$ homogeneous parts of $f_1$ and $g_1$, as explained below).

\noindent
\emph{Proof: } As $h_4\neq 0$, the degree restriction forces
$\deg{g_2} < n$. And since $\deg{f_2} < n$, we have the relation
$c_3[g_1]_n=-h_4[g_2]_{n-1}$. If $c_3\neq 0$, we have $[g_1]_n \in
(h_4)$, contradicting robustness of $[f_1]_n$ as then $[f_1]_n
= [g_1]_n + h_2[g_2]_{n-1} \in (h_2,h_4)$. {\tiny \qed}\\

\noindent
Therefore Equation~\ref{eqn:degres} gives,
$$
\insquar{\begin{array}{c} f_1\\f_2\end{array}} = \mat{1 & h_2\\0
  & c_4 + h_4}\insquar{\begin{array}{c} g_1\\g_2\end{array}}
$$
with $h_4\neq 0$. Also, since $[f_2]_{n+1} = [f_2]_n = 0$ this implies that $[g_2]_n = [g_2]_{n-1} = 0$. Hence, $[g_1]_n = [f_1]_n$ is $4$-robust. This argument can be extended now to $g_1$
and $g_2$. Notice that the degree of $g_1$ remains $n$. However, since there are only finitely many matrices in the sequence, there must come a point when this degree drops below $n$. At this point we get a contradiction as $[g_1]_n = 0$ (reusing symbol) which contradicts robustness.
\end{proof}

\noindent
We only need to finish the proof of Lemma~\ref{lem:degres}. 

\begin{proof}[Proof of Lemma~\ref{lem:degres}]
Suppose we have an equation of the form 
\begin{equation}\label{eqn:degreslem}
\insquar{\begin{array}{c} f_1\\f_2\end{array}} = \mat{h_1 + c_1 & h_2
  + c_2\\h_3 + c_3 & h_4 + c_4}\insquar{\begin{array}{c} g_1\\g_2\end{array}}
\end{equation}
On comparing degree $n+1$ terms, we have
\begin{eqnarray*}
h_1[g_1]_n + h_2[g_2]_n & = & 0\\
h_3[g_1]_n + h_4[g_2]_n & = & 0
\end{eqnarray*}
If $h_3$ and $h_4$ (a similar reasoning holds for $h_1$ and $h_2$)
were not proportional (i.e. not multiple of each other), then the above equation would imply $[g_1]_n,
[g_2]_n \in (h_3,h_4)$. Then, 
$$
[f_1]_n = h_1[g_1]_{n-1} +
h_2[g_2]_{n-1} + c_1[g_1]_n + c_2[g_2]_n \in (h_1,h_2,h_3,h_4)
$$
contradicting the robustness of $[f_1]_n$. Thus, $h_3$ and $h_4$ (as well as $h_1$ and $h_2$)
are proportional, in the same ratio as $[-g_2]_n$ and $[g_1]_n$. Using
an appropriate column operation, Equation~\ref{eqn:degreslem}
simplifies to
$$ \insquar{\begin{array}{c} f_1\\f_2\end{array}} = \mat{c_1 &
  h_2 + c_2\\c_3 & h_4 + c_4}\insquar{\begin{array}{c}
    g_1\\g_2\end{array}}
$$ If $c_1=0$, then together with $[g_2]_n = 0$ we get $[f_1]_n = h_2[g_2]_{n-1}$
contradicting robustness. Therefore $c_1 \neq 0$ and another column
transformation would get it to the form claimed.
\end{proof}

\section{Concluding remarks}

We give a new perspective to identity testing of depth $3$ arithmetic circuits by showing an 
equivalence to identity testing of depth $2$ circuits over $\mathsf{U}_2(\mathbb{F})$. 
The reduction implies that identity testing of a width-$2$ algebraic branching program 
is at least as hard as identity testing of depth $3$ circuits. 

We also give a deterministic 
polynomial time identity testing algorithm for depth $2$ circuits 
over any constant dimensional commutative algebra. Our algorithm crucially exploits 
an interesting structural result involving local rings. This naturally poses the 
following question - Can we use more algebraic insight on non-commutative algebras to 
solve the general problem? The solution for the commutative case does not seem to 
give any interesting insight into the non-commutative case. But we have a very specific 
non-commutative case at hand. The question is - Is it possible to use properties very specific 
to the ring of $2 \times 2$ matrices to solve identity testing for depth $3$ circuits?

\section*{Acknowledgement}
This work was started when the first author visited
Hausdorff Center for Mathematics, Bonn. We thank Marek
Karpinski for the generous hospitality and several discussions.
We also thank Manindra Agrawal for several insightful
discussions on this work. And finally thanks to V
Vinay for many useful comments on the first draft of this paper. 

\bibliographystyle{alpha}
\bibliography{Xbib}

\appendix
\section{Appendix} \label{sec:BC}

For the sake of completeness, we provide a proof of the result by
Ben-Or and Cleve~\cite{BC88}. 

\begin{theorem}\cite{BC88} \label{thm:BC}
  Let $E$ be an arithmetic formula of depth $d$ with fan-in (of every gate) bounded by
  $2$. Then, there exists a sequence of $3\times 3$ matrices, whose
  entries are either variables or constants, of length at most $4^d$
  such that one of the entries of their product is $E$.
\end{theorem}
\begin{proof}
  The proof is by induction on the structure of $E$. The base case when
  $E = c \cdot x_i$ is computed as,
  $$
  \matr{1 & & \\ & 1 & \\ c \cdot x_i & & 1}
  $$
  Suppose $E = f_1 + f_2$ and that we have inductively constructed
  sequences computing $f_1$ and $f_2$. Then the following equation gives a
  sequence for $E$. 
  $$ \matr{1 & & \\ & 1 & \\ f_1 & & 1}\matr{1 & & \\ & 1 & \\ f_2 & &
    1} = \matr{1 & & \\ & 1 & \\ f_1+f_2 & & 1}
  $$
  If $E = f_1\cdot f_2$, then the following sequence computes $E$
  $$ \matr{1 & & \\ -f_2 & 1 & \\ & & 1}\matr{1 & & \\ & 1 & \\ &f_1 &
    1}\matr{1 & & \\ f_2& 1 & \\ & & 1}\matr{1 & & \\ & 1 & \\ &-f_1 &
    1} = \matr{1 & & \\ & 1 & \\ f_1f_2 & & 1}
  $$
  Applying the above two equations inductively, it is clear that $E$ can be computed by a sequence of length at most $4^d$. 
\end{proof}

\end{document}

%% file: preamble.tex
\usepackage{latexsym}
\usepackage{amsmath}
\usepackage{amssymb}
\usepackage{amsthm}
\usepackage{graphicx}
\usepackage{hyperref}
\usepackage{algorithmic}
\usepackage{complexity}

\newtheorem{theorem}{Theorem}[section]
\newtheorem{theorem*}{Theorem}
\newtheorem{corollary}[theorem]{Corollary}
\newtheorem{lemma}[theorem]{Lemma}

\newtheorem{definition}[theorem]{Definition}

\newtheorem{problem}[theorem]{Problem}

\theoremstyle{definition}

\newcommand{\inparen}[1]{\left(#1\right)}             
\newcommand{\inbrace}[1]{\left\{#1\right\}}           
\newcommand{\insquar}[1]{\left[#1\right]}             

\newcommand{\F}{\mathbb{F}}

\renewcommand{\algorithmicrequire}{\textbf{Input:}}

%% file: matrix_PIT.bbl
\newcommand{\etalchar}[1]{$^{#1}$}
\begin{thebibliography}{ALM{\etalchar{+}}98}

\bibitem[AB99]{AB99}
Manindra Agrawal and Somenath Biswas.
\newblock {Primality and Identity Testing via Chinese Remaindering}.
\newblock In {\em FOCS}, pages 202--209, 1999.

\bibitem[Agr05]{A05}
Manindra Agrawal.
\newblock {P}roving {L}ower {B}ounds {V}ia {P}seudo-random {G}enerators.
\newblock In {\em FSTTCS}, pages 92--105, 2005.

\bibitem[AJMV98]{AJMV98}
Eric Allender, Jia Jiao, Meena Mahajan, and V.~Vinay.
\newblock {N}on-{C}ommutative {A}rithmetic {C}ircuits: {D}epth {R}eduction and
  {S}ize {L}ower {B}ounds.
\newblock {\em Theor. Comput. Sci.}, 209(1-2):47--86, 1998.

\bibitem[AKS04]{AKS04}
Manindra Agrawal, Neeraj Kayal, and Nitin Saxena.
\newblock {PRIMES} is in {P}.
\newblock {\em Ann. of Math}, 160(2):781--793, 2004.

\bibitem[ALM{\etalchar{+}}98]{ALMSS98}
Sanjeev Arora, Carsten Lund, Rajeev Motwani, Madhu Sudan, and Mario Szegedy.
\newblock {Proof Verification and the Hardness of Approximation Problems}.
\newblock {\em Journal of the ACM}, 45(3):501--555, 1998.

\bibitem[AMS08]{AMS08}
Vikraman Arvind, Partha Mukhopadhyay, and Srikanth Srinivasan.
\newblock New results on noncommutative and commutative polynomial identity
  testing.
\newblock In {\em IEEE Conference on Computational Complexity}, pages 268--279,
  2008.

\bibitem[AV08]{AV08}
Manindra Agrawal and V~Vinay.
\newblock {A}rithmetic circuits: {A} chasm at depth four.
\newblock In {\em FOCS}, pages 67--75, 2008.

\bibitem[BC88]{BC88}
Michael {Ben-Or} and Richard Cleve.
\newblock {Computing Algebraic Formulas Using a Constant Number of Registers}.
\newblock In {\em STOC}, pages 254--257, 1988.

\bibitem[BW05]{BW05}
Andrej Bogdanov and Hoeteck Wee.
\newblock More on noncommutative polynomial identity testing.
\newblock In {\em IEEE Conference on Computational Complexity}, pages 92--99,
  2005.

\bibitem[CDGK91]{CDGK91}
Michael Clausen, Andreas W.~M. Dress, Johannes Grabmeier, and Marek Karpinski.
\newblock {On Zero-Testing and Interpolation of k-Sparse Multivariate
  Polynomials Over Finite Fields}.
\newblock {\em Theor. Comput. Sci.}, 84(2):151--164, 1991.

\bibitem[CK97]{CK97}
Zhi-Zhong Chen and Ming-Yang Kao.
\newblock {Reducing Randomness via Irrational Numbers}.
\newblock In {\em STOC}, pages 200--209, 1997.

\bibitem[CS04]{CS04}
Steve Chien and Alistair Sinclair.
\newblock Algebras with polynomial identities and computing the determinant.
\newblock In {\em FOCS}, pages 352--361, 2004.

\bibitem[KI03]{KI03}
Valentine Kabanets and Russell Impagliazzo.
\newblock {D}erandomizing polynomial identity tests means proving circuit lower
  bounds.
\newblock In {\em STOC}, pages 355--364, 2003.

\bibitem[KS01]{KS01}
Adam Klivans and Daniel~A. Spielman.
\newblock {Randomness efficient identity testing of multivariate polynomials}.
\newblock In {\em STOC}, pages 216--223, 2001.

\bibitem[KS07]{KS07}
Neeraj Kayal and Nitin Saxena.
\newblock {P}olynomial {I}dentity {T}esting for {D}epth 3 {C}ircuits.
\newblock {\em Computational Complexity}, 16(2), 2007.

\bibitem[Lov79]{L79}
L{\'a}szl{\'o} Lov{\'a}sz.
\newblock {On determinants, matchings, and random algorithms}.
\newblock In {\em FCT}, pages 565--574, 1979.

\bibitem[LV98]{LV98}
Daniel Lewin and Salil~P. Vadhan.
\newblock {Checking Polynomial Identities over any Field: Towards a
  Derandomization?}
\newblock In {\em STOC}, pages 438--447, 1998.

\bibitem[Nis91]{Nis91}
Noam Nisan.
\newblock Lower bounds for non-commutative computation.
\newblock In {\em STOC}, pages 410--418, 1991.

\bibitem[RS04]{RS04}
Ran Raz and Amir Shpilka.
\newblock {D}eterministic {P}olynomial {I}dentity {T}esting in
  {N}on-{C}ommutative {M}odels.
\newblock In {\em IEEE Conference on Computational Complexity}, pages 215--222,
  2004.

\bibitem[Sch80]{S80}
Jacob~T. Schwartz.
\newblock {F}ast {P}robabilistic {A}lgorithms for {V}erification of
  {P}olynomial {I}dentities.
\newblock {\em J. ACM}, 27(4):701--717, 1980.

\bibitem[Sha90]{Shamir90}
Adi Shamir.
\newblock {IP=PSPACE}.
\newblock In {\em FOCS}, pages 11--15, 1990.

\bibitem[VSBR83]{VSBR83}
Leslie~G. Valiant, Sven Skyum, S.~Berkowitz, and Charles Rackoff.
\newblock Fast parallel computation of polynomials using few processors.
\newblock {\em SIAM J. Comput.}, 12(4):641--644, 1983.

\bibitem[Zip79]{Z79}
Richard Zippel.
\newblock {P}robabilistic algorithms for sparse polynomials.
\newblock {\em EUROSAM}, pages 216--226, 1979.

\end{thebibliography}
